\documentstyle[multicol,aps,prb,psfig]{revtex}

\begin{document}

 \draft

\title{Kondo and mixed valence regimes in multi-level quantum dots}

\author{A. L. Chudnovskiy and S. E. Ulloa \\
Department of Physics and Astronomy, and Condensed Matter and
Surface Sciences Program \\ Ohio University, Athens, Ohio
45701--2979}

\maketitle \centerline{(10 September 2000)}

\begin{abstract}
We investigate the dependence of the ground state of a multi-level
quantum dot on the coupling to an external fermionic system and on
the interactions in the dot. As the coupling to the external
system increases, the rearrangement of the effective energy levels
in the dot signals the transition from the Kondo regime to a mixed
valence (MV) regime. The MV regime in a two level dot is
characterized by an intrinsic mixing of the levels in the dot,
resulting in non-perturbative sub- and super-tunneling phenomena
that strongly influence the Kondo effect.
\end{abstract}

\pacs {1999 PACS: 72.15.Qm, 73.23.-b, 71.10.-w, 71.10.Hf}

\begin{multicols}{2}  \narrowtext

The Kondo effect arises from the coherent screening between a
localized spin and the spin of surrounding mobile electrons,
producing for example anomalous transport properties in metals
with magnetic impurities.\cite{Hewson}  Recently, however, there
has been a great deal of experimental activity in systems where
{\em an individual} localized spin is probed directly, either in
small metallic clusters, \cite{Ralph} in isolated magnetic atoms
on metallic surfaces, \cite{Crommie-Li} and in quantum dots
defined in semiconductor systems.
\cite{DavidGG,DGG,SC,JS,Simmel,Maurer} This activity has been
generated in great part by interesting and specific theoretical
predictions. \cite{Raikh,Lee,Wilkins,Meir}

A great deal of information has been obtained about the Kondo
effect in both the linear and non-linear transport regimes thanks
to experiments in and theory for quantum dots. Several conditions
have been analyzed, including large dots (with a quasi-continuous
energy spectrum), \cite{DavidGG,DGG,GHL} one-level dots (well
described by the Anderson impurity model),
\cite{Raikh,Lee,Wilkins,Meir,Phillips,Koenig} and the two- and
multi-level quantum dots with discrete energy spectrum.
\cite{Pohjola,Inoshita,Yeyati}  The ability to change the
structural parameters of the semiconductor quantum dot makes
these systems particularly attractive for the experimental
investigation of the Kondo effect.  This flexibility invites the
exploration of the transition from the Kondo to the non-Kondo
ground state of the dot as either the shape, coupling to the
external system or the gate voltage applied to the dot are
changed. For instance, experiments show different sequences of
Kondo and non-Kondo ground states in quantum dots, as the gate
voltage (hence the number of electrons in the dot) is varied.
\cite{DavidGG,DGG,SC,JS,Simmel,Maurer}  To address these
Kondo--non-Kondo transitions, we consider a model of a multilevel
quantum dot with discrete energy spectrum.  Our results are quite
interesting, as they describe a rich behavior for different
parameter values of experimental accessibility.  Previous
investigations on both single- and multi-level models have
uncovered interesting features of the Kondo effect, including
Kondo peaks in the conductance and the associated density of
states, their temperature dependence, and other features.
\cite{DGG,Simmel,Pohjola,Inoshita,Yeyati,20',20"}   However, the
typical approximation made in models of multilevel quantum dots
with discrete energy levels is to neglect the strong mixing
between the energy levels of the dot due to the interaction with
the external fermionic system. This approximation incorporates the
external fermionic system only as a broadening of the individual
levels in the quantum dot. \cite{Langreth1}

In this paper we show that the mixing among energy levels of the
dot due to the coupling to the external system leads to a
qualitative rearrangement of the levels at large coupling (in a
regime we call `mixed valence' (MV)).  This is characterized by
the appearance of effective dot levels exhibiting {\it sub-} and
{\it super-tunneling} to the leads.\cite{Shab-Ulloa}  In the
linear regime, we explore the alternation between spinless and
spinful ground states, as function of the interaction constants
in the dot and the coupling to the external fermionic system, as
the gate voltage (dot occupation) changes. Depending on the
various dot parameters, we find two regimes for {\em weak
coupling} to the leads: the ``Hund's rule'' regime, where the dot
tends to have maximal absolute value of spin, and successive
addition of electrons does not cancel the Kondo
effect;\cite{21',21"} and the even-odd Kondo--non-Kondo regime,
where the absolute spin of the dot is minimized, i.e., it is
nonzero only if the number of electrons in the dot is
odd.\cite{DavidGG}  For {\em strong coupling} with the leads, the
appearance of sub- and super-tunneling quasi-levels results in
changes of the occupation sequence of the dot in the MV regime,
in sharp contrast with the single-level quantum dot. This
behavior may yield the observed different sequences of Kondo and
non-Kondo ground states seen sometimes in experiments.
\cite{JS,Simmel,Maurer}

In what follows we consider explicitly a two-level quantum dot in
the linear regime at zero temperature, both for simplicity and
ease of presentation.  However, most of our description and
conclusions are valid for higher level-multiplicity.  Details will
be shown elsewhere. \cite{AC-SU}  The Hamiltonian of the model
can be written as
\begin{eqnarray}
H &=& \sum_{l,\sigma} \{E_l \hat{c}^+_{l,\sigma}\hat{c}_{l,\sigma}
+U_1\sum_{l\neq l'}\hat{n}_l\hat{n}_{l'}
-U_2(\vec{S}_l)^2\}+J\vec{S}_1 \cdot \vec{S}_2 \nonumber \\
 &+& \gamma\sum_{l,\sigma}(\hat{c}^+_{l,\sigma}\hat{a}_{0,\sigma}+
\hat{a}^+_{0,\sigma}\hat{c}_{l,\sigma}) +H_F(\hat{a}^+_{r,\sigma}
, \hat{a}_{r,\sigma}). \label{H}
\end{eqnarray}
Here, the fermionic operators $\hat{c}^+_{l,\sigma},
\hat{c}_{l,\sigma}$ describe the state (orbital) in the dot with
index $l=1,2$, and spin index $\sigma$.
$\hat{n}_{l,\sigma}=\hat{c}^+_{l,\sigma}\hat{c}_{l,\sigma}$ is the
particle number operator of the state $(l,\sigma)$ and
$\hat{n}_l=\hat{n}_{l\uparrow}+\hat{n}_{l\downarrow}$.
$\vec{S}_l=\hat{c}^+_{l,\alpha}{\bf \vec{\sigma}_{\alpha\beta}}
\hat{c}_{l,\beta}$ is the spin operator for level $l$. The
operators $\hat{a}^+_{r,\sigma}, \hat{a}_{r,\sigma}$ refer to the
fermions in the external electron system (the leads). There is a
tunnel coupling $\gamma$ between the dot and the external system
at point $r=0$, and $H_F$ is the Hamiltonian of the external
fermionic system.  This one-lead geometry is schematically
depicted in the inset of Fig.\ 1.

The basic interactions in the quantum dot are the Hubbard-like
repulsion between the electrons on a given level of the dot,
represented here by the constant $U_2$, and the density-density
repulsion between the charges on {\em different} energy levels via
the interaction constant $U_1$. We also include an exchange
interaction between the spins of the two levels via the
interaction constant $J$. Typically, the range of $J$ is $0\leq
|J|\leq U_1$.  In what follows we consider the antiferromagnetic
exchange ($J>0$), the role of the ferromagnetic exchange being
less interesting. The various constants of these interactions
depend mostly on the overlap integrals between the wave functions
of the different energy levels. Therefore, they can be changed by
changing the shape of the quantum dot, which is experimentally
accessible. \cite{DavidGG,Maurer}

To deal with the external fermionic system, we note that due to
the $\delta(r)$-like interaction with the dot, one can consider
only the s-wave scattering of the external electrons on the dot,
and hence use a one-dimensional description of the external
system. \cite{Affleck}  We use a discretized representation of
the external electron system, so that the resulting problem
reduces to the coupling of the quantum dot states to a 1D
fermionic chain $H_F = t\sum_{i=1}^L (\hat{a}^+_{i\sigma}
\hat{a}_{i+1,\sigma} + \hat{a}^+_{i+1,\sigma}\hat{a}_{i\sigma})$
(similar to the linear chain form for the s-d model in Ref.\
[\onlinecite{Hewson}]).

Integrating the fermionic degrees of freedom consecutively,
starting from the site directly coupled to the quantum dot, we
find that the influence of the external fermionic system on the
quantum dot renormalizes the coupling constants in the system,
and most importantly, gives rise to an additional term of the form
$i\kappa \sum_{l,l'} \hat{c}^{\omega_n +}_{l\sigma}
\hat{c}_{l'\sigma}^{\omega_n}$ in the effective Hamiltonian,
where $\omega_n$ denotes the Matsubara frequency. Consecutive
elimination of external degrees of freedom yields recurrent
relations for the running interaction constants; we find
 \begin{eqnarray}
 \tilde{\omega}_{j+1} &=& \omega _n + \tilde{\omega}_{j}
 t^2 /D_{nj} \nonumber \\
 \tilde{\gamma}_{j+1} &=& - \tilde{\gamma}_{j}
 t^2 /D_{nj}  \nonumber \\
 \tilde{\kappa}_{j+1} &=& \tilde{\kappa}_j + i \omega_n
 \tilde{\gamma}_{j}^2 /D_{nj}  \, ,
\end{eqnarray}
where $D_{nj} = {\omega_n \tilde{\omega}_{j} + t^2} $, and $j$ is
the integration step index.

These relations can be directly iterated. As we are interested in
the interaction of the dot with infinite leads, we look for fixed
points of the recurrent relations.  This yields an effective dot
coupling for small Matsubara frequencies given by $ \kappa\approx
(\gamma^2 / 2t) {\rm sgn} (\omega_n)$, where the hopping in the
fermionic chain $t$ is of the order of the bandwidth of the
external fermionic system. \cite{Hewson,Affleck} While being
approximately constant at small frequencies, the mixing $\kappa$
decays at large frequencies. This property insures the transition
from the mixed valence regime for shallow energy levels, to the
Kondo regime for deep levels.

The state of the quantum dot is characterized by the occupation
of different energy levels. To determine the ground state of the
dot, we treat the many particle interactions by means of the
Hubbard -- Stratonovich decoupling of the four-fermionic terms
with subsequent mean field (Hartree) approximation for the
bosonic decoupling fields. \cite{AC-SU}  At the mean field level,
the spin and occupation of a given level of the dot are
characterized by the average values of the decoupling fields
$\vec{M}_l$ (the average spin of the level $l=1$ or 2 -- assumed
frozen along the $z$-direction), and $R_l$ (the average
charge/number of electrons on level $l$). The mean field solutions
for the isolated quantum dot, $\kappa=0$, allow us to classify the
states of the quantum dot as follows: i) $M_l=0$, $R_l=0$: the
level $l$ is empty; ii) $M_l=\pm 1$, $R_l=1$: the level $l$ is
singly occupied, spin $1/2$; iii) $M_l=0$, $R_l=2$: the level $l$
is doubly occupied, spin $0$. Further, we denote the states of the
quantum dot as $(n,m)$ with $n$ and $m$ being the occupation of
the first and the second energy levels, respectively.

In the mean field treatment, we replace the two level quantum dot
with interactions by a non-interacting quantum dot with
quasi-energy levels that depend on the ground state of the quantum
dot. This approach allows one to describe correctly the properties
of the quantum dot in the ground state and its gapless
excitations, including the Kondo effect, which is the aim of our
treatment. At the same time, the quasi-energy levels do not
reproduce the gapped excitation spectrum of the quantum dot
measured, for example, in optical experiments.

For the {\em isolated} quantum dot, the positions of the
quasi-energy levels are given by
$\epsilon^{\downarrow,\uparrow}_1 = E_1+R_2 U_1\pm
[M_1(2U_2+J)-M_2 J]$, $\epsilon^{\downarrow,\uparrow}_2 = E_2+R_1
U_1\pm [M_2(2U_2+J)-M_1 J]$, where the sign ``$+$'' (``$-$'')
corresponds to $\downarrow$ ($\uparrow$) spin.  As the gate
voltage lowers, more electrons occupy the quantum dot. Whereas
the spin of states with 0, 1, 3 and 4 electrons in the dot is
defined uniquely, the state with 2 electrons in the dot can have
total spin 1, if the electrons occupy different energy levels (the
state (1,1)), or total spin 0 in the states (0,2) or (2,0). Which
of these states is chosen depends on the relation between the
interaction constants in the dot. Taking the degenerate case
$E_1=E_2$ and comparing the free energies of different states of
the isolated dot with total occupation 2, we find, that the state
(1,1) is realized under the condition $U_1<2U_2+3J$, i.e., if the
Hubbard and exchange interactions that favor the magnetic state
dominate over the density-density repulsion between the electrons
on different levels. In the opposite case, the nonmagnetic (0,2)
or (2,0) states of the quantum dot are realized for two electrons
in the dot. Extending this result for the multilevel dot with
almost degenerate single particle levels, the condition
$U_1<2U_2+3J$ corresponds to the Hund's rule regime, hence a
sequence of consecutive Kondo states as the dot charge increases.
The opposite condition corresponds to the alternation of the
Kondo and non-Kondo ground states for odd and even numbers of
electrons, respectively, as one expects for separate single-level
dots.

The coupling to the leads via a nonzero coupling $\kappa$ results
in the broadening and mixing of the energy levels that are given
now by the expression $ z^\sigma_\pm=\left(\epsilon^\sigma_1
+\epsilon^\sigma_2\pm \sqrt{(\epsilon^\sigma_1 -
\epsilon^\sigma_2)^2 - 4\kappa^2}\right)/2-i\kappa $. In the
limit of zero level mixing (zero tunneling), the values of
$z^\sigma_\pm$ coincide with $\epsilon^\sigma_{1,2}$. Small level
mixing $\kappa$ leads to small deviations of the solutions from
the above mentioned values, although the structure of the
quasi-energy spectrum remains the same.

In contrast, in the case of strong level mixing, when the
condition $2\kappa > |\epsilon_1-\epsilon_2|$ is satisfied, the
quasi-energy spectrum changes {\em qualitatively}, reflecting the
transition into the regime we call mixed-valence (MV). This regime
is realized if the level spacing in the noninteracting dot is
small. Then the square root in the expression for the
$z_\pm^\sigma$ turns out to be purely imaginary. The positions of
the energy levels for a given spin direction coincide. It is
convenient to rearrange the energies $z_{\pm}$ introducing the
values $z_1$ and $z_2$ in the following manner $z_1\equiv
z_+\theta(\omega_n)+z_-\theta(-\omega_n)$, $z_2\equiv
z_+\theta(-\omega_n)+z_-\theta(\omega_n)$. Then we obtain $
z_{1,2} =\left(\epsilon_1 +\epsilon_2\right)/2-
i\left(\kappa\mp\sqrt{\kappa^2- (\epsilon_1
-\epsilon_2)^2/4}\right) {\rm sgn} (\omega_n) \label{z12} $. In
the limit of large $\kappa$ one has two degenerate levels for each
spin projection, one strongly broadened ($z_2$) and the other one
with strongly suppressed broadening ($z_1$).

Since in the MV regime the quasi-levels are degenerate (even in
the presence of interactions), the electrons fill the dot
pairwise, each pair containing two electrons with the {\it same}
projection of the spin. (One could say that the strong mixing
provides an effective ferromagnetic interaction between spins.)
The sequence of ground states obtained by changing the gate
voltage is modified correspondingly. For example, neglecting the
exchange interaction, and in the regime $E_1-E_2>2U_2-U_1>0$, the
sequence of ground states in the dot changes from ``empty dot
$\rightarrow$ lower level singly occupied (Kondo state)
$\rightarrow$ lower level doubly occupied $\rightarrow$ upper
level singly occupied with lower level full (Kondo state)
$\rightarrow$ both levels doubly occupied´´ in the weak coupling
regime [or $(0,0)\rightarrow (1,0) \rightarrow (2,0) \rightarrow
(2,1) \rightarrow (2,2)$ in our $(n,m)$ notation], to ``empty dot
$\rightarrow$ both levels singly occupied (Kondo state with spin
1) $\rightarrow$ both levels doubly occupied´´ in the mixed
valence regime [or $(0,0) \rightarrow (1,1) \rightarrow (2,2)$].
The first sequence results in two Kondo peaks in the conductance
(odd occupation) separated by two non-Kondo states (even
occupation), whereas in the MV case one can see only one Kondo
triplet-like peak (in the state with both levels singly
occupied). This difference in ground state sequences will clearly
have experimentally observable consequences.
\cite{DavidGG,DGG,SC,JS,Simmel,Maurer}  Notice, however, that the
contribution to the Kondo effect from the two levels differs
qualitatively. Whereas the broad supertunneling level results in
behavior typical of the one-level dot in the MV regime (where
charge fluctuations are strong, and one observes strong
temperature dependence of the Kondo peak \cite{DGG}), the narrow
subtunneling level contributes as a deep Kondo level in a
one-level quantum dot (with large spin fluctuations and universal
temperature dependence).

Note that the sub- and super-tunneling rearrangement of the
energy levels and related changes in the state of the quantum dot
described here can only occur in the multilevel (two-level here)
dot.  Moreover, this can be calculated theoretically only if one
takes fully into account the effect of {\it mixing} of energy
levels in the dot through the coupling to the external fermionic
system.  This non-perturbative effect is not present in other
treatments, either because the strong coupling is explicitly
neglected, \cite{Yeyati} or because it is treated only
perturbatively. \cite{Pohjola}

\begin{figure}[p]
\psfig{file=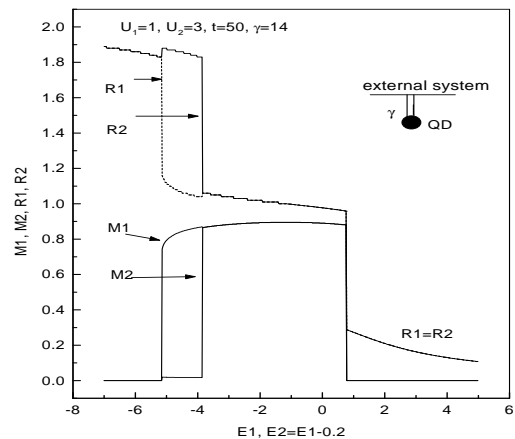,width=10cm,height=10cm,angle=0}
\vspace{-15ex}
 \caption{State
of the quantum dot with changing gate voltage. The quantum dot is
in the MV regime for $E_1$ close to $E_F=0$. For the deep energy
levels, in the region where levels 1 and 2 are in different states
($M_1\neq 0$, $M_2=0$), the quantum dot is in the Kondo regime.
The transition between the MV and the Kondo regime coincides with
level 2 jumping from $M_2=M_1$ to $M_2=0$ (at $E_1 \approx
-3.9$).} \label{mv-kon}
\end{figure}

A typical change of the state of the dot with gate voltage is
shown in Fig.\ \ref{mv-kon}. Close to the Fermi level ($E_F=0$,
for $E_1, E_2 \approx 0.84, 0.64$) the spin and the occupation of
both levels jump simultaneously, the values of the average spin
and charge on both levels being equal, $M_1=M_2$, $R_1=R_2$. The
dot is, therefore, in the MV regime. As the gate voltage lowers
the levels in the dot, the MV regime breaks down (at $E_1\approx
-3.9$), the level 2 becomes doubly occupied and nonmagnetic
($R_2\approx 2$, $M_2=0$), whereas the level 1 remains singly
occupied. The quantum dot is in the Kondo regime. Finally, the
level 1 becomes doubly occupied also (at $E_1\lesssim -5.1$). The
dot is completely filled. Note, that in the nonmagnetic states of
the dot, the spin is identically zero, whereas the average charge
assumes a continuously varying nonzero value due to the coupling
to the external system. The different behavior of the spin and
charge of the dot is similar to the quantization of the spin and
the absence of quantization of the charge in a quantum dot,
predicted recently in the limit of a large dot. \cite{GHL}

\begin{figure}[p]
\psfig{file=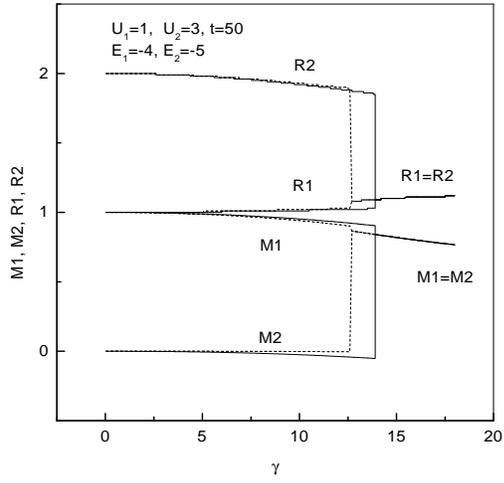,width=12cm,height=15cm,angle=0}
 \vspace{-35ex}
 \caption{State of the quantum dot with changing tunneling amplitude
$\gamma$ -- transition from weak coupling regime to MV regime.
$J=1$ -- solid lines; $J=0$ -- dashed lines. Nonzero
antiferromagnetic exchange shifts the transition to the MV regime
to larger coupling $\gamma$. } \label{mv-afm}
\end{figure}

The exchange interaction generally favors the magnetic (1,1) state
of the quantum dot with both levels singly occupied. However,
whereas the ferromagnetic exchange does not affect the transition
between the Kondo and MV regimes, the AFM exchange suppresses the
MV regime, effectively increasing the interlevel separation
$\epsilon^\sigma_1-\epsilon^\sigma_2$. At the same time, the
Kondo effect is suppressed by the antiferromagnetic correlation
of spins in the dot. \cite{Lee,Phillips}  The influence of the
AFM exchange interaction on the MV regime is illustrated in Fig.\
\ref{mv-afm}. Here the change of the ground state of the dot is
shown versus the tunneling amplitude $\gamma$ for zero and
nonzero values of the AFM coupling $J$. We can see that at the
transition to the MV regime the charge state of the dot changes
from (1,2) (Kondo state with three electrons in the dot, spin
1/2) to (1,1) (Kondo state with two electrons in the dot, spin 1).
Therefore, increasing the coupling to the external fermionic
system causes the delocalization of one of the electrons in the
quantum dot. The nonzero AFM exchange $J$ shifts the transition
to the higher values of the coupling $\gamma$ thus competing with
the MV regime.

Different ground state behavior has been identified in a
two-level quantum dot (although most of our conclusions can be
generalized to the multi-level case).  This behavior depends on
interaction parameters and coupling to the external leads,
resulting in a situation not described before: a strong-coupling
`mixed valence' regime, which produces Kondo-like behavior for
even numbers of electrons in the dot.  One can anticipate that if
a sample exhibits consecutive Kondo peaks for both odd and even
occupation, the signature of the MV regime would be a
non-universal temperature dependence of the conductance, and its
low-temperature unitary limit would clearly exceed the
single-channel level.

This work was supported in part by US Department of Energy Grant
No.\ DE-FG02-91ER45334.

\vspace{-50ex}

\vspace{-8ex}

\end{multicols}
\end{document}